\documentclass{aa}
\usepackage{graphicx}
\usepackage{natbib}
\usepackage{amsmath}
\bibpunct{(}{)}{;}{a}{}{,}

\newcommand{\ffh}{\widehat f}
\newcommand{\gh}{\widehat g}
\newcommand{\Kh}{\widehat K}

\newcommand{\Kmh}{\widehat {\cal{K}}}

\newcommand{\deltah}{\widehat{\delta}}

\newcommand{\varsigmah}{\widehat{\varsigma}}

\newcommand{\fbh}{\boldsymbol {\widehat{f}}}

\newcommand{\wh}{\widehat w}

\newcommand{\fb}{\boldsymbol{f}}

\newcommand{\Lb}{\boldsymbol{L}}

\newcommand{\Hb}{\boldsymbol{H}}
\newcommand{\Ib}{\boldsymbol{I}}
\newcommand{\gb}{\boldsymbol{g}}
\newcommand{\gbt}{\boldsymbol{\tilde g}}
\newcommand{\gt}{{\tilde g}}
\newcommand{\fbt}{\boldsymbol{\tilde f}}

\newcommand{\ab}{\boldsymbol{a}}
\newcommand{\eb}{\boldsymbol{e}}
\newcommand{\rb}{\boldsymbol{r}}
\newcommand{\wb}{\boldsymbol{w}}

\newcommand{\varthetab}{\boldsymbol{\vartheta}}
\newcommand{\varsigmab}{\boldsymbol{\varsigma}}
\newcommand{\Acb}{\boldsymbol{\cal A}}

\newcommand{\Ab}{\boldsymbol{A}}

\newcommand{\Pb}{\boldsymbol{P}}
\newcommand{\varrhob}{\boldsymbol{\varrho}}

\newcommand{\nmat}{{\cal N}}
\newcommand{\dmat}{{\cal D}}

\begin{document}

\title{Estimation of Regularization Parameters \\ in Multiple-Image Deblurring}

   \author{R. Vio\inst{1}
          \and
          P. Ma\inst{2}
          \and
          W. Zhong\inst{3}
          \and
          J. Nagy\inst{4}
          \and
          L. Tenorio\inst{5}
          \and
          W. Wamsteker\inst{6}
          }

   \offprints{R. Vio}

   \institute{Chip Computers Consulting s.r.l., Viale Don L.~Sturzo 82,
              S. Liberale di Marcon, 30020 Venice, Italy\\
              ESA-VILSPA, Apartado 50727, 28080 Madrid, Spain\\
              \email{robertovio@tin.it}
         \and
		  Department of Statistics, Harvard University, Cambridge, MA 02138, USA\\
		  \email{pingma@stat.harvard.edu}
         \and
		  Department of Statistics, Purdue University, West Lafayette, IN 47907, USA \\
		  \email{wenxuan@stat.purdue.edu}
         \and
              Department of Mathematics and Computer Science, Emory University, Atlanta, GA 30322, USA. \\
              \email{nagy@mathcs.emory.edu}
         \and
              Department of Mathematical and Computer Sciences, Colorado School of Mines, Golden CO 80401, USA \\
              \email{ltenorio@Mines.EDU}
        \and
             ESA-VILSPA, Apartado 50727, 28080 Madrid, Spain\\
             \email{willem.wamsteker@esa.int}
             }

\date{Received .............; accepted ................}

\abstract{We consider the estimation of the regularization parameter for the simultaneous
deblurring of multiple noisy images via Tikhonov regularization. We approach the problem in three ways. We first
reduce the problem to a single-image deblurring for which the regularization parameter can be estimated
through a classic generalized cross-validation (${\rm GCV}$) method. A modification of this
function is used for correcting the undersmoothing typical of the original technique.
With a second method, we minimize an average least-squares fit to the images and define a new ${\rm GCV}$ function.
In the last approach, we use the classical ${\rm GCV}$ on a single higher-dimensional image obtained by concatanating
all the images into a single vector.
With a reliable estimator of the regularization parameter, one can fully exploit the
excellent computational characteristics typical of direct deblurring methods, which, especially for
large images, makes them competitive with the more flexible but much slower iterative
algorithms. The performance of the techniques is analyzed through numerical experiments.
We find that under the independent homoscedastic and Gaussian assumptions made on the noise,
the three approaches provide almost identical results
with the first single image providing the practical advantage that no new software is required and the same image can
be used with other deblurring algorithms.
}
\titlerunning{${\rm GCV}$ for multiple image deblurring}
\authorrunning{R. Vio, P. Ma, W. Zhong, J. Nagy, L. Tenorio, \& W. Wamsteker}
\maketitle

\section{Introduction}

An important problem in image processing is the
simultaneous deblurring of a set of observed images (of a fixed object),
each of which is degraded by a different point spread function (PSF).
Consider, for example, images
obtained by the Large Binocular Telescope (LBT). This instrument consists of two $8.4 {\rm m}$ mirrors on
a common mount with a spacing of $14.4 {\rm m}$ between the centers \citep{ang98}. For a given orientation of
the telescope, the diffraction-limited resolution along the center-to-center baseline is equivalent to that
of a $22.8 {\rm m}$ mirror, while the resolution along the perpendicular direction is that of a single $8 {\rm m}$
mirror. One way to obtain an image with improved and uniform spatial resolution is to simultaneously
deconvolve the images taken with different orientations of the telescope.

In a recent paper, \citet{vio04} presented an efficient solution based on a single image
that improves the performance of iterative algorithms developed in the context of a
least-squares approach \citep{ber00}.
An important advantage of iterative algorithms is the
ease with which constraints such as nonnegativity of the solution can be implemented.
However, this incurs a high computational cost (especially in case of slow
convergence). In addition, there is a lack of a reliable stopping criteria. Although these issues
may not be critical
for images of moderate size, they are critical for large images ($\approx 10^7-10^8$ pixels) or
studies that require the deblurring of a large number of images. In this respect, direct methods are potentially
more useful. For example, with a fixed regularization parameter, Tikhonov deblurring
requires only two two-dimensional discrete Fourier transforms.

A limitation of any regularization method, and in particular of Tikhonov's approach, is the need to get
a computationally efficient and reliable estimate of the regularization parameter.
We present three different methods for such a task in the case of multiple spatially invariant PSFs and
additive
Gaussian noise. In Section~\ref{sec:formalization}, we formalize the problem and propose three solutions
in Sect.~\ref{sec:single}. Their performance is studied through numerical experiments in Sect.~\ref{sec:experiments}.
Section~\ref{sec:conclusions} closes with final comments.

\section{Formalization of the problem} \label{sec:formalization}
The image restoration problem is to find an estimate $f(x,y)$ of a fixed two-dimensional object $f_0(x,y)$
given $p$ observed images $g_j(x,y)$ ($j=1,2, \ldots, p$)
each of which is degraded by a different spatially invariant blurring operator; that is,
\begin{equation} \label{eq:model1}
g_j(x,y) = \iint K_j (x-x', y-y') f_0(x',y') dx' dy',
\end{equation}
where each $K_j(x,y)$ is a known space invariant PSF.

In practice, the observed images are discrete $N \times N$ arrays of pixels
(images are assumed square for simplicity) contaminated by noise. Therefore, model~(\ref{eq:model1})
has to be recast in the form
\begin{equation} \label{eq:model2}
g_j(m,n) = \sum_{m',n'=0}^{N-1} K_j(m-m', n-n') f_0(m',n') + w_j(m,n),
\end{equation}
where $w_j(m,n)$ will be assumed to be Gaussian white noise with standard deviation $\sigma_{w_j} = \sigma_w$.
Note however that the methods in Sect.~\ref{sec:single} may be used when the noise is not white but has a correlation
structure that is known up to a constant factor.

If the central peak (i.e., support) of each PSF is much smaller than the image and
the object does not have structures near the image boundaries, then the convolution
product in Eq.~(\ref{eq:model2}) can be well approximated by cyclic convolution, which
leads to
\begin{equation} \label{eq:model}
\gh_j(m,n) = \Kh_j(m,n) \ffh_0(m,n) + \wh_j(m,n),
\end{equation}
where the symbol ``$~\widehat{}~$'' denotes the discrete Fourier transform (DFT).

In order to estimate $f_0$ using Tikhonov regularization, it is helpful to use matrix-vector notation.
Rewrite Eq.~(\ref{eq:model2}) as
\begin{equation} \label{eq:modela}
\gb_j = \Ab_j \fb_0 + \wb_j,
\end{equation}
where the arrays $\gb_j$, $\fb_0$ and $\wb_j$ are obtained by column stacking $g_j(m,n)$,
$f_0(m,n)$, and $w_j(m,n)$, respectively. The matrix $\Ab_j$ is block-circulant and defined
by cyclic convolution with the $j$th PSF.

\citet{ber00} estimate $\fb_0$ by Tikhonov regularization where the estimate
$\fb_\lambda$ minimizes
an average fit to the images subject to a smoothness constraint:
\begin{eqnarray} \label{eq:Tikmin1}
\fb_{\lambda} & = &\arg\min\,\left(\,\sum_{j=1}^{p} \| \Ab_j \fb - \gb_j \|^2 + \lambda^2 \|\,\Lb\fb\,\|^2\,\right)\\
& = &\arg\min\,\left(\,\| \Acb \fb - \varrhob \|^2 + \lambda^2 \|\,\Lb\fb\,\|^2\,\right)\label{eq:Tikmin1b},
\end{eqnarray}
where the global imaging matrix and global image are defined, respectively, as
\begin{equation}\label{eq:global}
\Acb =
\begin{pmatrix}
       \Ab_1\\
       \Ab_2\\
       \vdots\\
       \Ab_p
\end{pmatrix},\quad \varrhob =
\begin{pmatrix}
       \gb_1\\
       \gb_2\\
       \vdots\\
       \gb_p
\end{pmatrix},
\end{equation}
$\lambda$ is the regularization parameter and $\Lb$ is, usually, the identity matrix (for energy bound) or
a discrete Laplacian operator (for smoothness constraint) \citep{wahba90}.
The solution $\fb_{\lambda}$ satisfies the normal equations
\begin{equation} \label{eq:single}
\left(\,\Ab + \lambda^2 \Lb^T\Lb\,\right)\fb_{\lambda} = \varthetab,
\end{equation}
where
\begin{equation}\label{eq:AandG}
\Ab = \sum_{j=1}^{p} \Ab_j^T\Ab_j,\qquad \varthetab = \sum_{j=1}^{p} \Ab^T_j\gb_j.
\end{equation}
For a fixed value of $\lambda$, its explicit solution is given by:
\begin{equation} \label{eq:explicit}
\ffh_{\lambda}(m,n) = \sum_{j=1}^p \frac{\Kh_j^*(m,n) \gh_j(m,n)}{\Kmh(m,n) + \lambda^2},
\end{equation}
where $\Kmh(m,n) = \sum_{j=1}^p |\Kh_j(m,n)|^2$.

In the rest of the paper we consider the selection of the regularization parameter $\lambda$ for
three different methods whose normal equations reduce in each case to \eqref{eq:single}.

\section{Estimating the regularization parameter} \label{sec:single}

\subsection{A first single-image approach}
The Tikhonov regularization \eqref{eq:Tikmin1} is transformed into a single-image problem \eqref{eq:Tikmin1b} through the
use of the higher dimensional global image $\varrhob$. We now define an alternative single-image
approach whose dimensionality is the same as that of the original images.

The solution of the normal equations \eqref{eq:single} depends on the original data $\varrhob$
only through $\varthetab$.
In fact, in the case of homoscedastic uncorrelated Gaussian noise, $\varthetab$ and $\varrhob$ contain the
same information about $\fb_0$. This can be justified with a sufficiency argument on $\varthetab$ \citep{leh98}.
It thus seems natural to pose the problem directly through $\varthetab$: Since the expected value of $\varthetab$ is
$\Ab\fb_0$, we propose to estimate $\fb_0$ by regularizing the misfit
$ \Vert\, \Ab \fb - \varthetab \,\Vert^2$. But one problem with $\varthetab$ is that its covariance matrix is
${\rm Cov}(\varthetab) = \sigma_w^2\,\Ab$, which is no longer proportional to the identity and therefore
the classical generalized cross-validation (${\rm GCV}$) is not expected to work well.
In fact, our simulations show that ${\rm GCV}$
consistently underestimates $\lambda$ when used with $\varthetab$.
However, since $\Ab$ is known, one can easily
correct for this correlation using a square-root of the symmetric matrix $\Ab$: the image
\begin{equation}
\varsigmab = \Ab^{-1/2}\varthetab
\end{equation}
has mean $\Ab^{1/2}\fb_0$ and covariance
matrix $\sigma_w^2\,\Ab^{-1/2}\Ab\Ab^{-1/2} = \sigma_w^2\Ib$. Hence, we define the Tikhonov functional
\begin{equation} \label{eq:tikhonovv}
\fb_{\lambda} = {\rm argmin}\left(\, \Vert\, \Ab^{1/2} \fb - \varsigmab\,\Vert^2 + \lambda^2 \Vert\,
\Lb \fb \,\Vert^2\, \right).
\end{equation}
It can be easily shown that the normal equations for \eqref{eq:tikhonovv} are given by \eqref{eq:single}.
This result indicates that the functional~(\ref{eq:Tikmin1}) can be expressed in terms
of a single-image $\varsigmab$ with PSF $\Ab^{1/2}$, whose
Fourier domain representation is
\begin{equation} \label{eq:varsig}
\varsigmah(m,n) = \sum_{j=1}^p \frac{\Kh_j^*(m,n) \gh_j(m,n)}{\Kmh^{1/2}(m,n)},
\end{equation}
and $\Kmh^{1/2}(m,n)$, respectively. Through $\varsigmah(m,n)$, the solution~(\ref{eq:explicit}) for
$\ffh_{\lambda}(m,n)$ can be expressed in the form
\begin{equation}
\ffh_{\lambda}(m,n) = \frac{\Kmh^{1/2}(m,n) \varsigmah(m,n)}{\Kmh(m,n) + \lambda^2}.
\end{equation}
Since the covariance matrix of $\varsigmab$ is proportional
to the identity, we can make
use of the classical generalized cross-validation (${\rm GCV}$) function
to estimate $\lambda$ \citep[e.g., ][]{wahba90,vio03}:
\begin{equation}
{\rm GCV}^{(\varsigmab)}(\lambda) = \frac{1}{N^2}\frac{\|\,\Ab^{1/2}\fb_{\lambda}-\varsigmab\,\|^2}
{\left(\,1-{\rm Tr}[\Hb^{(\varsigmab)}]/N^2\,\right)^2},
\end{equation}
where $\Hb^{(\varsigmab)}= \Ab^{1/2}\left(\,\Ab + \lambda^2 \Lb^t\Lb\,\right)^{-1}\Ab^{1/2}$,
with frequency domain representation
\begin{align} \label{eq:gcvo}
 & {\rm GCV}^{(\varsigmab)}(\lambda)   \nonumber  \\ =
 &                N^2 \sum_{m,n}
                 \Bigg| \frac{\varsigmah(m,n) |\deltah(m,n)|^2}
                       {\Kmh(m,n)+\lambda^2 |\deltah(m,n)|^2}
                 \Bigg|^2 \nonumber \\
                   /
  &                \left(
                         \sum_{m,n}
                         \frac{|\deltah(m,n)|^2}
                              {\Kmh(m,n)+\lambda^2 |\deltah(m,n)|^2}
                  \right)^2.
\end{align}
Here, $\deltah(m,n)$ provides the two-dimensional DFT of matrix $\Lb$.
However, despite its theoretical justification and adequate practical performance, ${\rm GCV}$ tends to underestimate
the regularization parameter (too small a $\lambda$) in about $10\%$ of the cases. A corrected ${\rm GCV}$ can be used
to control this without sacrificing its generally good performance. The corrected weighted ${\rm GCV}$ is:
\begin{align}  \label{eq:gcvov}
  & {\rm GCV_{\alpha}}^{(\varsigmab)}(\lambda) \nonumber \\
                 & = \frac{1}{N^2}
                 \sum_{m,n}
                 \Bigg|\frac{\lambda^{2}\varsigmah(m,n)|\deltah(m,n)|^2}
                       {\Kmh(m,n)+\lambda^{2}|\deltah(m,n)|^2}
                 \Bigg|^{2} / \nonumber \\
                   & \left((1-\alpha)
                       + \frac{\alpha}{N^2}\sum_{m,n}
                         \frac{\lambda^{2}|\deltah(m,n)|^2}
                              {\Kmh(m,n)+\lambda^2|\deltah(m,n)|^2}
                   \right)^2
\end{align}
where  $\alpha > 1$. It reduces to (\ref{eq:gcvo}) when $\alpha = 1$ and yields
smoother estimates as $\alpha$ increases.

Empirical studies suggest good values of $\alpha$ for practical use in the range of 1.2 and 1.4.
Modifications of this sort have been suggested by various researchers \citep[e.g., ][]{nychka98}. However,
only recently \citep{cfn01, kg03} have systematic simulation studies been conducted to evaluate the
performance of the corrected ${\rm GCV}$ in the context of statistical nonparametric modeling.
The optimality of the corrected  ${\rm GCV}$ (\ref{eq:gcvov}) can be
interpreted in a way similar to that of ${\rm GCV}$ (\ref{eq:gcvo}).
Instead of estimating a relative loss function, Eq.~(\ref{eq:gcvov}) estimates a weighted combination of the average
squared bias and the average variance \citep{cfn01}. It should also be noted that ${\rm GCV}$ is a function of both
$\lambda$ and $\alpha$, but it is difficult to decide whether ${\rm GCV}$ should be minimized over $\alpha$. Current
research efforts are still trying to resolve this issue.

It is important to emphasize that the problem of underestimation is not unique to ${\rm GCV}$,
it is common to virtually all
regularization parameter selection techniques. For example, the generalized maximum likelihood (GML)
estimate of $\lambda$ \citep{wahba85} asymptotically tends to deliver slightly smaller estimates than ${\rm GCV}$.
An attractive feature of our simple correction is that it is not restricted to ${\rm GCV}$,
and sheds a fresh light on
underestimation corrections for GML and other approaches.

The minimization of the ${\rm GCV}$ function cannot be done analytically, some numerical
iteration method has to be used. In our simulation we use the Newton-Raphson algorithm.

\subsection{Multiple-image approach}

Although the single-image $\varsigmab$ is adequate
for applications where the most important information is preserved in the single image,
here we define ${\rm GCV}$ for functions~(\ref{eq:Tikmin1}) and \eqref{eq:Tikmin1b}, and
compare their performances with that obtained with the single-image $\varsigmab$.

We start by defining a multiple-image ordinary
cross validation (${\rm CV}$) function. One such function can be obtained
by estimating the prediction error of each image
using a leave-one-out estimate
\begin{equation}\label{eq:multicv}
{\rm CV}(\lambda)= \frac{1}{p}\sum_{j=1}^p \frac{1}{N^2}\sum_{i=1}^{N^2} (g_{ji}-\tilde{g}_{j,-i})^2.
\end{equation}
Here, $\tilde{g}_{j,-i}$ is the prediction based on the estimate of $\fb_0$ that uses
all the data except the $i$th entry in $\gb_j$.
Simple algebra leads to an equation for $\tilde{g}_{j,-i}$ in terms of $\gbt_j$ (see Appendix \ref{sec:gupdate})
\begin{equation}\label{eq:gupdate}
\tilde{g}_{j,-i} = g_{ji} + \frac{r_{j,i}}{1-(\Hb_j)_{ii}},\quad r_{j,i} = \tilde{g}_{ji} - {g}_{ji},
\end{equation}
where,
\begin{equation}\label{eq:gandH}
\gbt_{j} = \Ab_j \fb_\lambda,\quad \Hb_j = \Ab_j (\,\Ab + \lambda^2 \Lb^T\Lb\,)^{-1}\Ab_j^T.
\end{equation}
The multiple-image ${\rm CV}$ function can then be written as
\begin{equation}
{\rm CV}(\lambda)= \frac{1}{N^2\,p}\sum_{j=1}^p\sum_{i=1}^{N^2} \frac{r_{j,i}^2}{(1-(\Hb_j)_{ii})^2}.
\end{equation}
The multiple-image ${\rm GCV}$ is obtained by modifying \eqref{eq:multicv} to be rotationaly invariant.
That is, the value of the GCV function should
remain unchanged if each $\gb_i$ is rotated. There is no unique way to achieve such rotational invariance but one method
is to substitute the average diagonal value ${\rm Tr}\,[\Hb_j]/N^2$ for
$(\Hb_j)_{ii}$
\begin{equation} \label{eq:multigcv}
{\rm GCV}^{(\gb)}(\lambda)=\frac{1}{N^2\,p}\sum_{j=1}^p \frac{\|\,\rb_j\,\|^2}{(1-{\rm Tr}\,\Hb_j/N^2)^2},
\end{equation}
where $\rb_{j} = \gbt_{j} - {\gb}_{j}$.
Note that Eq.~(\ref{eq:multigcv}) reduces to the classical ${\rm GCV}$ when $p=1$.

The implementation of Eq.~(\ref{eq:multigcv}) is done in the frequency domain. It can be rewritten as
\begin{equation} \label{eq:gcv2}
{\rm GCV}^{(\gb)}(\lambda) = \frac{N^2}{p} \sum_{j=1}^p \frac{\nmat_j}{\dmat_j}
\end{equation}
where
\begin{equation} \label{eq:gcv2a}
\nmat_j=\sum_{m,n} \Bigg| \frac{\Kh_j(m,n) \gh(m,n)}{\Kmh(m,n) +
\lambda^2 |\deltah(m,n)|^2} - \gh_j(m,n) \Bigg|^2,
\end{equation}
with $\gh(m,n) = \sum_{l=1}^p \Kh^*_l(m,n) \gh_l(m,n)$, and
\begin{equation} \label{eq:gcv2b}
\dmat_j= \left[ \sum_{m,n} \left(1- \frac{|\Kh_j(m,n)|^2}{\Kmh(m,n) +\lambda^2 |\deltah(m,n)|^2}\right)\right]^2.
\end{equation}

To avoid underestimating the regularization parameter, the corrected ${\rm GCV}$ for (\ref{eq:multigcv}) can be derived as
\begin{equation} \label{eq:cmultigcv}
{\rm GCV}_{\alpha}^{(\gb)}(\lambda)=\frac{1}{N^2\,p}\sum_{j=1}^p \frac{\|\,\rb_j\,\|^2}{(1-\alpha~{\rm Tr}[\Hb_j]/N^2)^2}.
\end{equation}
Its form in the frequency domain is given by Eqs.~(\ref{eq:gcv2})-(\ref{eq:gcv2a}), and
\begin{equation} \label{eq:gcv2c}
\dmat_j= \left[ \sum_{m,n}\left( 1- \frac{\alpha |\Kh_j(m,n)|^2}{\Kmh(m,n) +\lambda^2 |\deltah(m,n)|^2}\right)\right]^2.
\end{equation}

\subsection{A second single-image approach}
An alternative form for the ${\rm GCV}$ functional~(\ref{eq:multigcv})
can be obtained by applying the standard
${\rm GCV}$ directly to Eq.~(\ref{eq:Tikmin1b}) \footnote{We thank Prof. M. Bertero for this suggestion.}
based on the single global-image $\varrhob$.
In this case the rotation invariance is enforced in a higher dimensional
space of concatenated images which leads to
\begin{eqnarray} \label{eq:cmultigcva}
{\rm GCV}_{\alpha}^{(\varrhob)}(\lambda) & = & \frac{1}{p N^2} \frac{\sum_j \Vert \,\rb_j\,\Vert^2}
{(\,1- \alpha ~{\rm Tr}[\Hb^{(\varrhob)}]/(p N^2)\,)^2} \nonumber \\
& = & \frac{1}{p N^2} \frac{\sum_j \Vert \,\rb_j\,\Vert^2}{(\,1-
\alpha ~\sum_j {\rm Tr}[\Hb_j]/(p N^2)\,)^2},
\end{eqnarray}
where $\Hb^{(\varrhob)} = \Acb(\Ab + \lambda^2 \Lb^t\Lb)^{-1}\Acb^t$.
%and
%\begin{equation} \label{eq:gcv3}
%{\rm GCV}^{(\varrhob)}(\lambda) = N^2  \frac{\sum_j \nmat_j}{\sum_j\dmat_j}.
%\end{equation}

Following the usual interpretation of the denominator in the ${\rm GCV}$ function \citep{wahba90}, we
see that ${\rm GCV}^{(\gb)}(\lambda)$ and ${\rm GCV}^{(\varrhob)}(\lambda)$ differ in that the former
weights each residual error by the noise degrees of freedom of each image; images with a larger
trace of $\Hb_j$ receive a higher weight.
In our case the two ${\rm GCV}$ functions provide identical results (see below).

It can also be shown that even though $\varsigmab$ and $\varrhob$ lead to the same normal equations,
their solutions may differ because their ${\rm GCV}$ may give different answers. In fact, one can check that
\begin{align}
&{\rm GCV}^{(\varsigmab)}(\lambda) - p\, {\rm GCV}^{(\varrhob)}(\lambda)  =  \nonumber \\
&\frac{1}{N^2}\sum_{j,k} \gb_j^t\Ab_j\Ab^{-1}\Ab_k^t\gb_k
-  \frac{1}{N^2}\sum_j \|\,\gb_j\,\|^2 .
\end{align}

\section{Some numerical experiments} \label{sec:experiments}

We present the results of numerical simulations to test the
reliability of the regularization parameter estimated through the ${\rm GCV}$ functions defined in
Sect.~\ref{sec:single}.

Figures~\ref{fig:ping_sat_10_1}-\ref{fig:ping_star_100_2} compare the deblurring of
an extended object with a sharp outline and a set of point-like objects.
We have chosen these particular examples because their restoration is a difficult problem.
Eight images are available for each object. In each case the PSF is a bidimensional Gaussian with
dispersion along the major axis set
to twelve pixels, and to four pixels along the minor axis. Their inclinations take equispaced
values in the range $0^{\circ} - 160^{\circ}$. Gaussian white noise is added with two different
levels of noise by setting the standard deviations to
$1 \%$ and $10 \%$, respectively, of the maximum value of the blurred images.
Three deblurring methods have been used: (I) single-image Tikhonov based on the functional~(\ref{eq:tikhonovv})
and the ${\rm GCV_{\alpha}^{(\varsigmab)}}$ function~(\ref{eq:gcvov}) ($\alpha=1.4$);
(II) Tikhonov based on the functional~({\ref{eq:Tikmin1}) and  ${\rm GCV_{\alpha}^{(\gb)}}$ function~(\ref{eq:cmultigcv})
($\alpha=1.4$); (III) a classic iterative {\it Projected Landweber Method}
applied to the single-image images for the iterative methods obtained as explained in \citet{vio04},
that can be used to enforce nonnegativity of the solution (a maximum number
of $2000$ iterations has been adopted, that is sufficient to provide results that do not
show further appreciable visual changes).

There is no need to test the ${\rm GCV}$ functional~(\ref{eq:cmultigcva}) because in our case
it is identical to~(\ref{eq:cmultigcv}). The reason is that the shape of all
the selected PSFs is the same, they only differ in their orientation, and since orientation
information is lost through the trace, the denominator $\dmat_j$ does not depend
on the image index $j$. That is, all the images share the same effective degrees of freedom.

Since the Tikhonov technique does not enforce the nonnegativity constraint, the deblurred images obtained
with the single-image approach are also shown with their negative values set to zero.
Finally, for each of the experiments, the true ``optimal'' value (in the sense of the minimization of the rms of
the true residuals)
for the regularization parameter provided by the ${\rm GCV}$ function~(\ref{eq:gcvov}) has been determined,
with an accuracy of two significant digits, by a simple grid search.
In all cases, the ``true'' and the estimated regularization parameter have been almost identical.
In particular, the relative difference between the rms of the true residuals
of the corresponding solutions was always less than $0.3 \%$.

As shown by the figures, the multiple and single-image approaches lead to very similar, if not identical,
residual rms values.

As expected, Projected Landweber provides the best results in terms of visual appearance and residual rms.
In fact, as is well known, the nonnegativity constraint limits the ringing effect
around the sharp structures in the image. However, in these experiments the difference seems important
only for point-like objects. In any case, these results are obtained at a very high
computational cost (each iteration requires the computation of two two-dimensional DFTs).

Apart from the nonnegativity constraint, the worst performance of the Tikhonov method is due to
the functional $\Lb$ that tends to smooth out structures with sharp features.
This characteristic makes Tikhonov more suited for the restoration of images of diffuse objects.
In fact, Figs.~\ref{fig:ping_galaxyD10_1}-\ref{fig:ping_galaxy100_2} show that for diffuse objects,
single-image Tikhonov provides competitive restoration quality \citep[see also ][]{ber98}. Because of the
presence of features close to the borders of the images, calculations for the image $\varsigma(m,n)$ and
the deblurring operation have been implemented with the windowing procedure as explained in \citet{vio04}.
This has required the removal of a border $50$ pixels thick from the image and has only
been done with the single-image Tikhonov approach because, given the necessity
of windowing each single image, it would be tortuous to implement the other version of the algorithm.

In order to test the stability of the $\lambda$ estimates, for each experiment $100$ realizations of the
corresponding noise process are generated and added to the image. In general, the estimates of this parameter
appear quite stable; as measured by the rms of the estimated values divided by their median, they fall
in the range $1\% - 2\%$. Fig.~\ref{fig:fig_hist} shows the histogram of the values of $\lambda$ obtained with
the single-image Tikhonov approach in the experiments with the worst relative variation.

\section{Concluding remarks} \label{sec:conclusions}

We have considered the estimation of the regularization parameter for the simultaneous Tikhonov deblurring
of multiple noisy images using three types of ${\rm GCV}$ functions based on different data transformations.
The first approach reduces the multiple-image problem to single-image deblurring that
%that is
%equivalent to a weighted Tikhonov regularization that penalizes deviations
%from the normal equations of the average least-squares fit. The weighting
allows the use of the classical
${\rm GCV}$, but we use a corrected ${\rm GCV}$ that reduces its tendency to undersmooth the final solution.
A second approach is perhaps the most natural as it is a straightforward extension of a classical single-image
approach. It aims to achieve a balance between the smoothness of the unknown image and its average goodness
of fit to the observed images. The regularization parameter is estimated through a multiple-image extension
of the traditional ${\rm GCV}$ function. The third approach is based on a higher dimensional
single image defined by concatenating all the images into a single vector and then using the classical
${\rm GCV}$.

Our numerical experiments indicate that the three approaches provide very similar
results, with the lower-dimensional single image providing
the practical advantage that no new software is required and the same image can
be used with other deblurring algorithms.

We have assumed that the noise in all the images is independent Gaussian and of constant variance.
Combinations of the methods we propose can be used when some of these assumptions are not met.
For example, if there are $k$ groups of images with different variances,
a weighted Tikhonov regularization method can be applied to a
multiple-image deblurring of $k$ single images. In additon, the multiple-image
approach can be used if the noise in the original image is correlated but the covariance
matrix is known up to a constant \citep[e.g., ][]{gu02}.

The extreme computational efficiency, coupled with the stability and reliabilty of the estimated regularization
parameters, make Tikhonov's method competitive with the more flexible,
but much slower, iterative algorithms. In fact, and especially for very large images,
even in situations where these algorithms are expected to provide better results (e.g., when the
nonnegativity constraint can be exploited), the Tikhonov approach is still a valuable resource
\citep[see also ][]{ber00}. For example, it can be used to obtain initial solutions for
iterative algorithms that may speed up the convergence. Figure~\ref{fig:fig_iter} shows the residual
rms as a function of iteration number for the Projected Landweber algorithm started at
a Tikhonov solution and at the uniform image traditionally used. The figure shows that
the saving in computation time can be considerable (here it is more than a factor of $10$).
\begin{figure}
        \resizebox{\hsize}{!}{\includegraphics{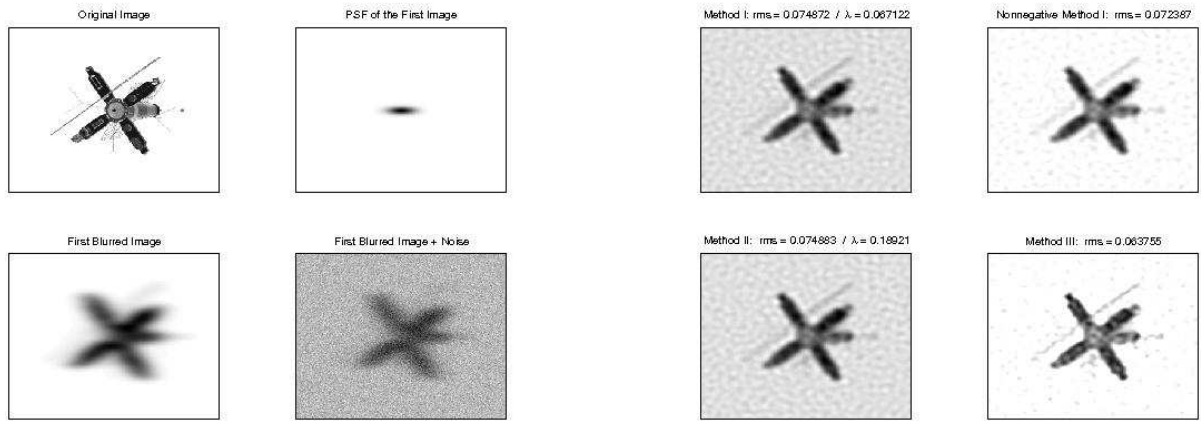}}
        \caption{Original image, blurred version and
        blurred + noise version of the first image in the set (see text) and corresponding PSF. The
        images are $256 \times 256$ pixels, and the standard deviation of the noise is $10\%$ of
        the maximum value of the blurred image.}
        \label{fig:ping_sat_10_1}
\end{figure}
\begin{figure}
        \resizebox{\hsize}{!}{\includegraphics{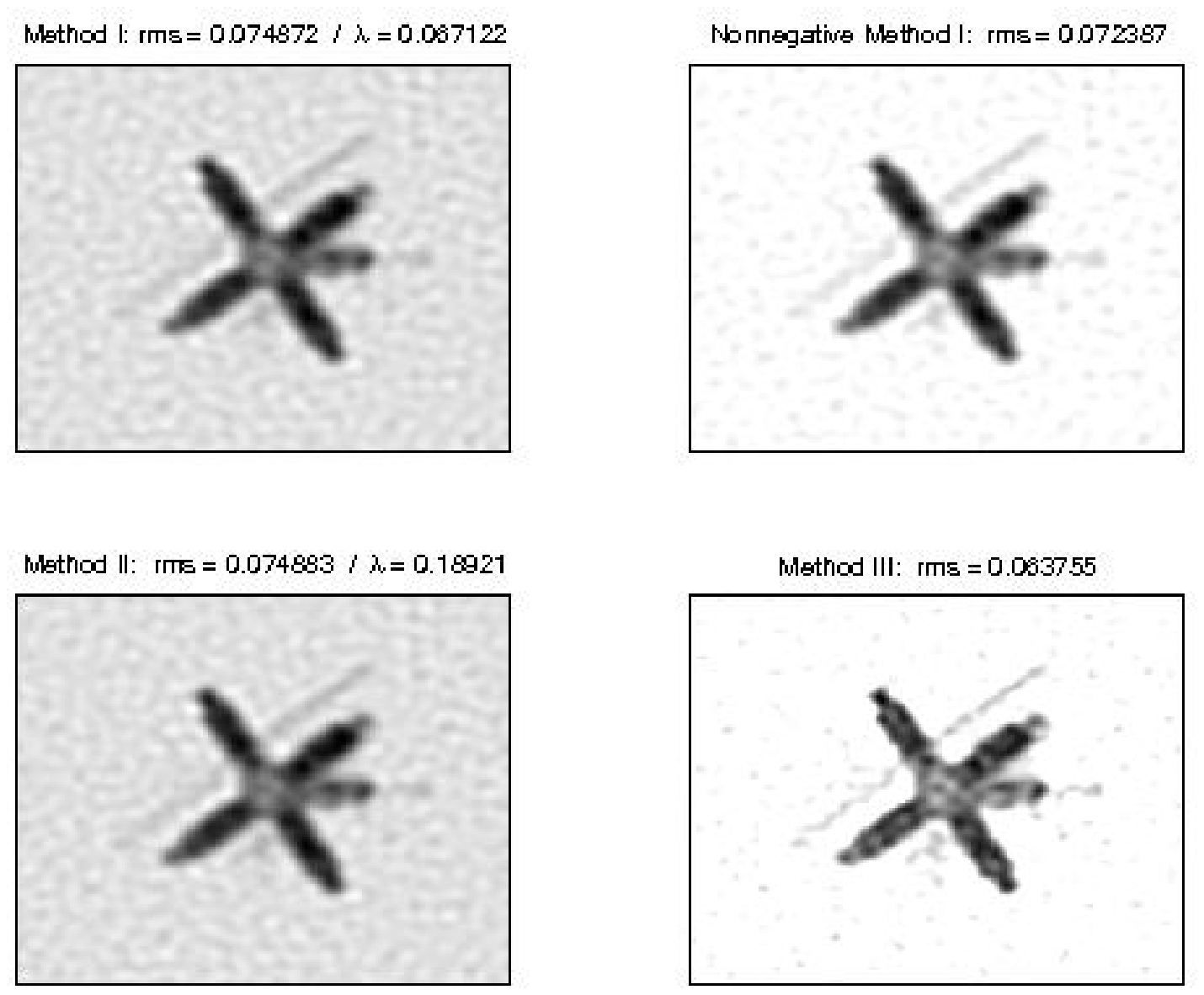}}
        \caption{Deblurring of the image in Fig.~\ref{fig:ping_sat_10_1} with the methods decribed in the text:
        (I) single-image Tikhonov~(\ref{eq:tikhonovv}) and ${\rm GCV^{(\varsigmab)}_{\alpha}}$ ~(\ref{eq:gcvov});
        its version with the negative elements zeroed;
        (II) Tikhonov~(\ref{eq:Tikmin1}) and ${\rm GCV^{(\gb)}_{\alpha}}$~(\ref{eq:cmultigcv});
        (III) a classic iterative Projected Landweber Method (convergence after $61$ iterations).
        Tikhonov has been used with $\Lb = \Ib$ and $\alpha = 1.4$.}
        \label{fig:ping_sat_10_2}
\end{figure}
\begin{figure}
        \resizebox{\hsize}{!}{\includegraphics{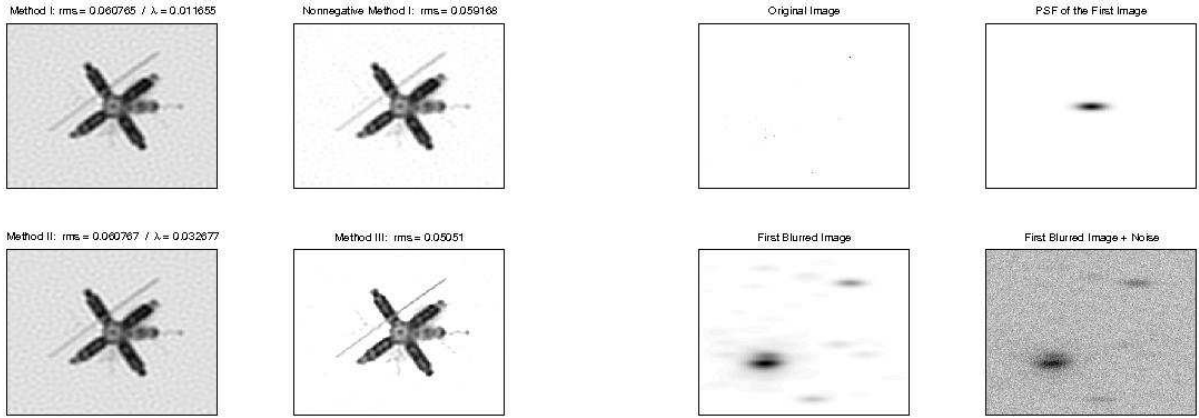}}
        \caption{The same as Fig.~\ref{fig:ping_sat_10_2} but obtained from an image contaminated
	  with noise whose standard deviation is $1\%$ of the maximum value of the blurred image.
        For the Projected Landweber algorithm $2000$ iterations have been carried out.}
        \label{fig:ping_sat_100_2}
\end{figure}
\begin{figure}
        \resizebox{\hsize}{!}{\includegraphics{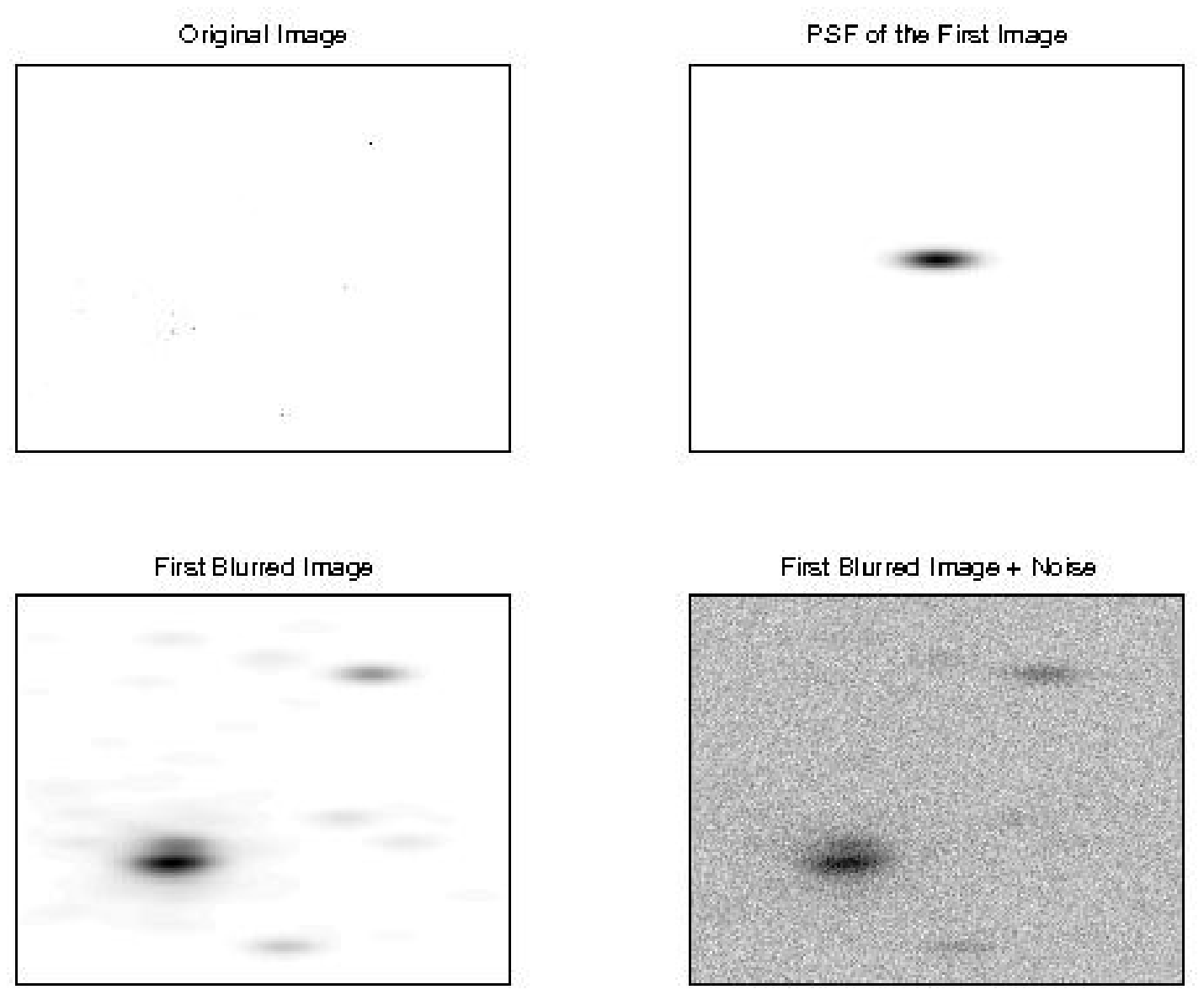}}
        \caption{Original image, blurred version and
        blurred + noise version of the first image in the set (see text) and corresponding PSF. The
        images are $256 \times 256$ pixels, and the standard deviation of the noise is $10\%$ of
        the maximum value of the blurred image.}
        \label{fig:ping_star_10_1}
\end{figure}
\begin{figure}
        \resizebox{\hsize}{!}{\includegraphics{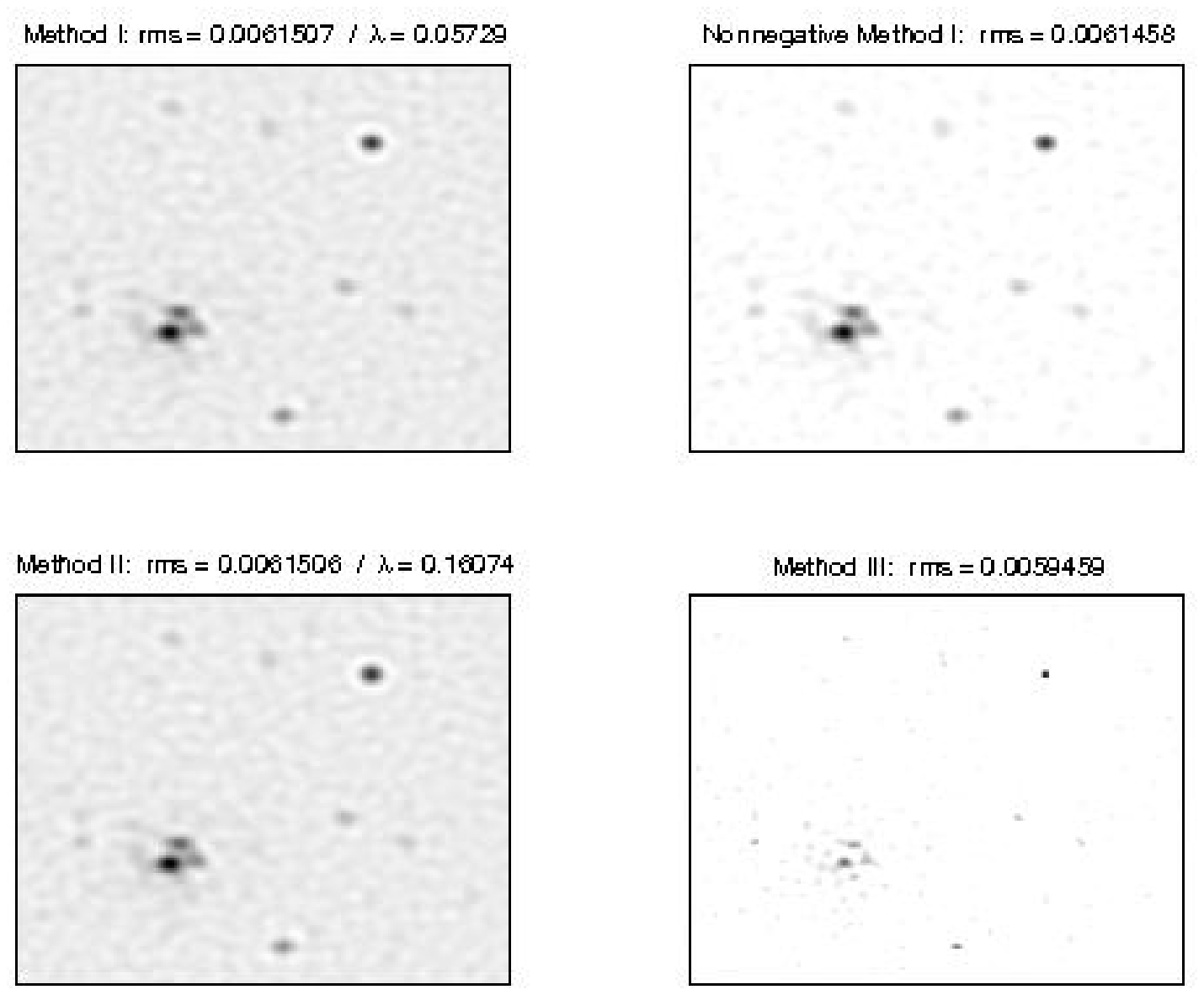}}
        \caption{Deblurring of the image in Fig.~\ref{fig:ping_star_10_1} with the methods described in the text:
        (I) single-image Tikhonov~(\ref{eq:tikhonovv}) and ${\rm GCV^{(\varsigmab)}_{\alpha}}$ ~(\ref{eq:gcvov});
        its version with the negative elements zeroed;
        (II) Tikhonov~(\ref{eq:Tikmin1}) and ${\rm GCV^{(\gb)}_{\alpha}}$~(\ref{eq:cmultigcv});
        (III) a classic iterative Projected Landweber Method ($2000$ iterations).
        Tikhonov has been used with $\Lb = \Ib$ and $\alpha = 1.4$.}
        \label{fig:ping_star_10_2}
\end{figure}
\begin{figure}
        \resizebox{\hsize}{!}{\includegraphics{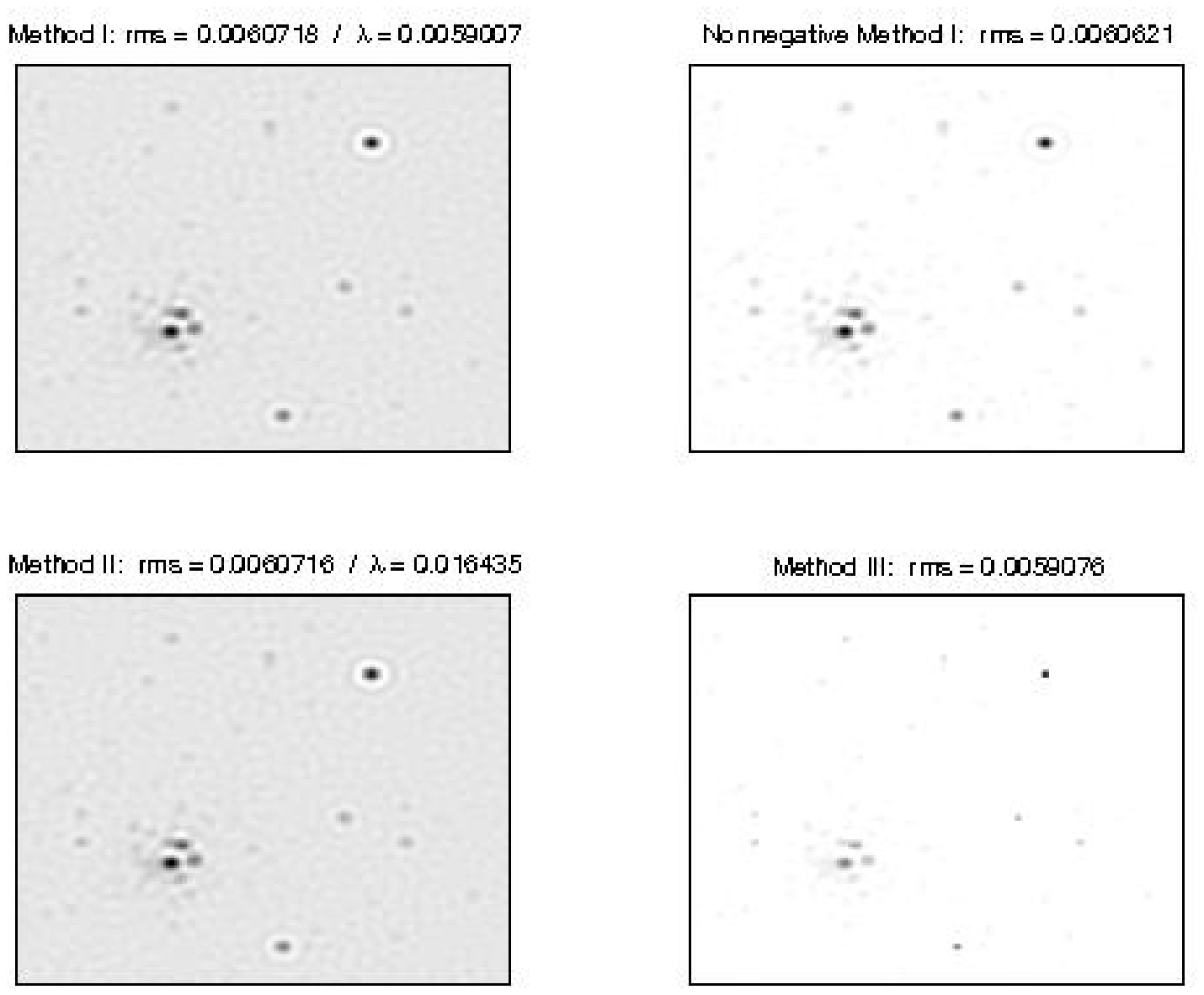}}
        \caption{The same as Fig.~\ref{fig:ping_star_10_2} but obtained from an image contaminated
	  with noise whose standard deviation is $1\%$ of the maximum value of the blurred image.}
        \label{fig:ping_star_100_2}
\end{figure}
\begin{figure}
        \resizebox{\hsize}{!}{\includegraphics{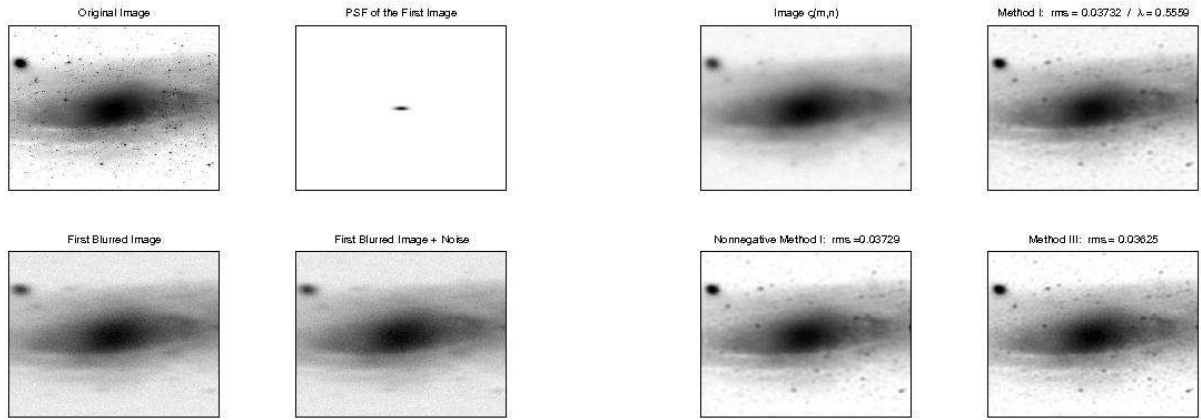}}
        \caption{Original image, blurred version and
        blurred + noise version of the first image in the set (see text) and corresponding PSF. The original
        images was $600 \times 600$ pixels, but only the central $500 \times 500$ pixels are displayed (see
        text). The standard deviation of the noise is $10\%$ of the standard deviation of the blurred image.}
        \label{fig:ping_galaxyD10_1}
\end{figure}
\begin{figure}
        \resizebox{\hsize}{!}{\includegraphics{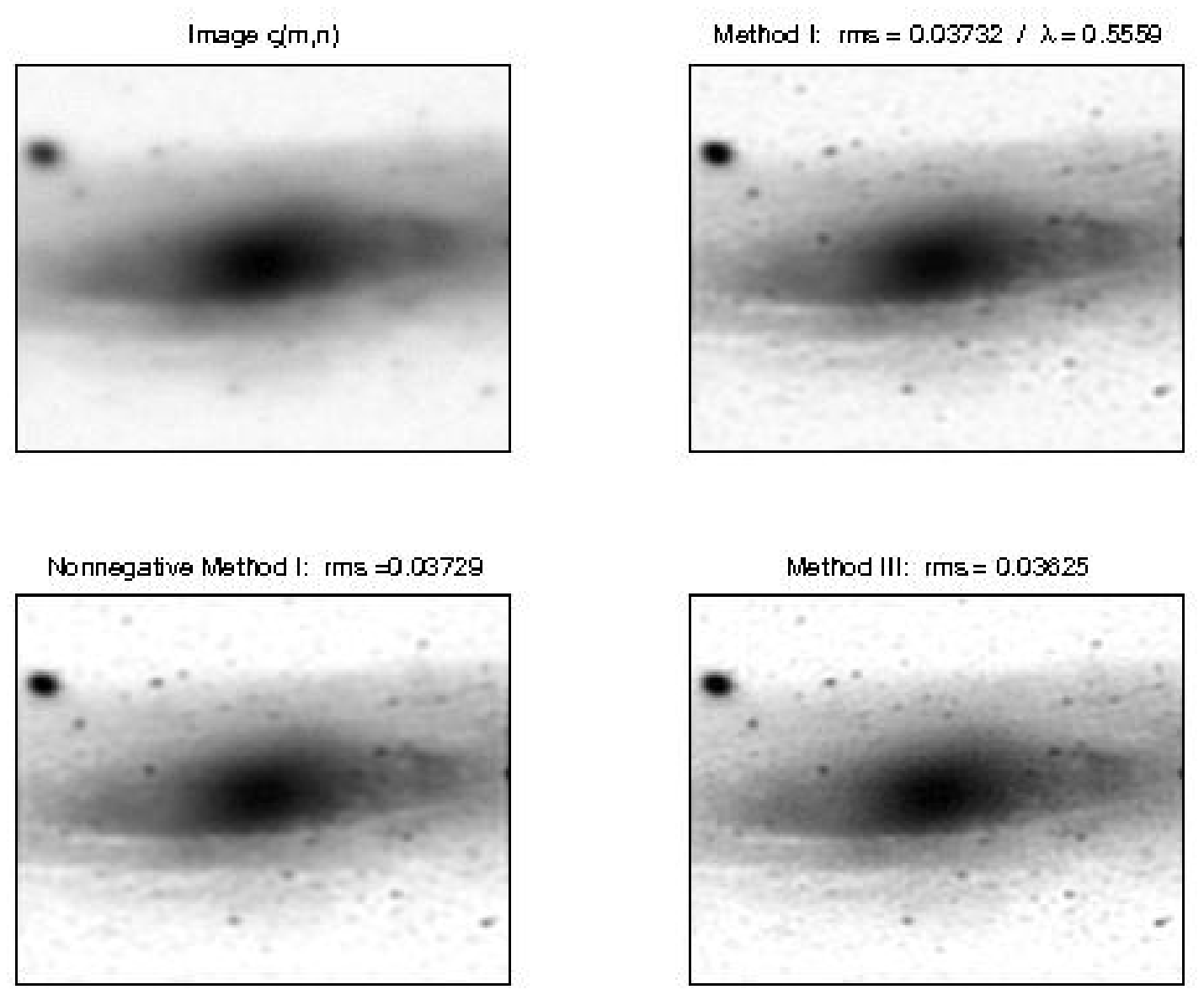}}
        \caption{Image $\varsigma(m,n)$ for the image shown in
	  Fig.~\ref{fig:ping_star_10_1}, single-image Tikhonov deblurring
	  ($\Lb = $ discrete Laplacian), its version with the negative elements zeroed, and
        the iterative Projected Landweber deblurring (convergence after $24$ iterations).}
        \label{fig:ping_galaxyD10_2}
\end{figure}
\begin{figure}
        \resizebox{\hsize}{!}{\includegraphics{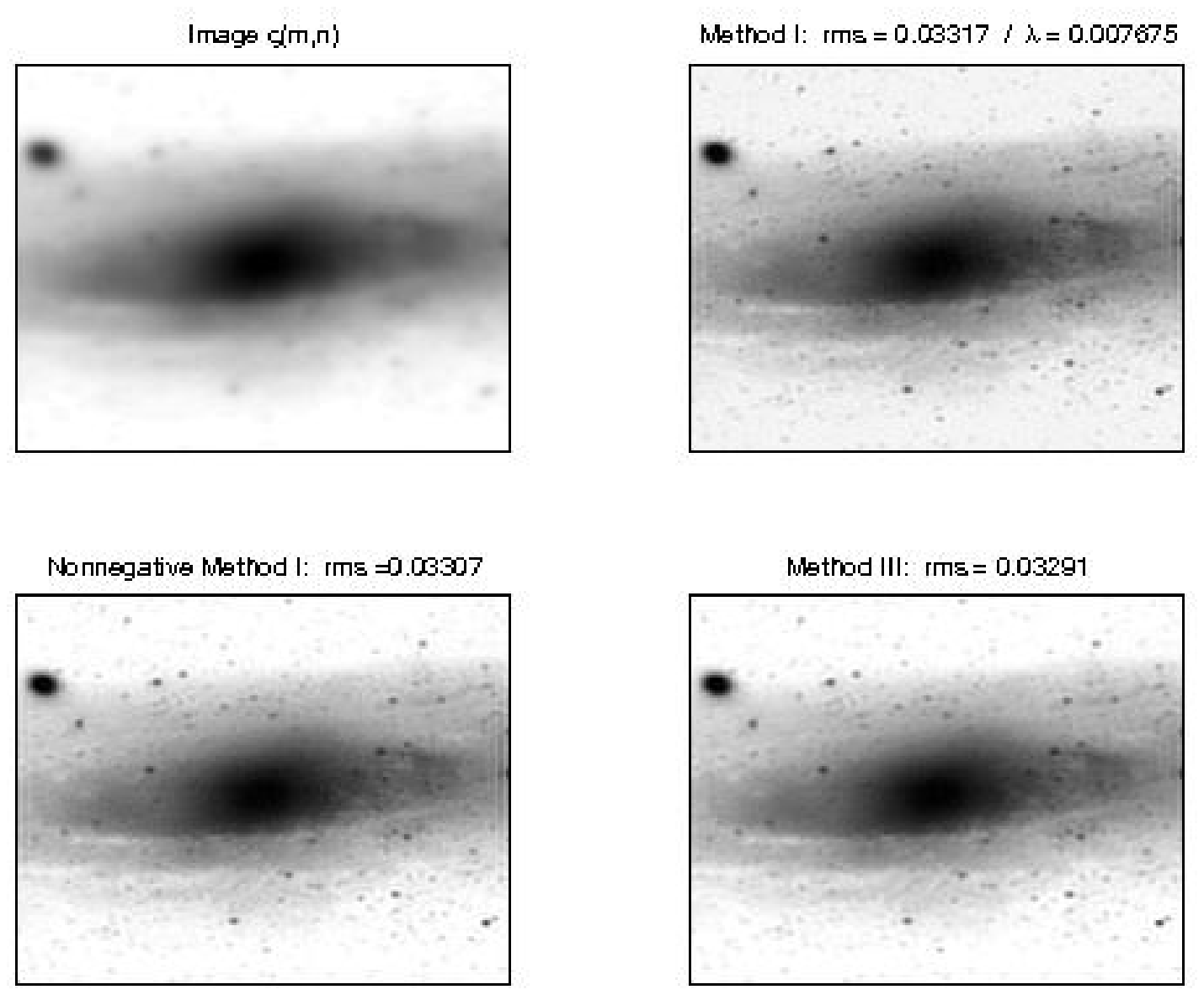}}
        \caption{The same as Fig.~\ref{fig:ping_galaxyD10_2} but obtained from an image contaminated
	  with noise whose standard deviation is $1\%$ of the standard deviation of the blurred image.
        Tikhonov has been used with $\Lb = \Ib$. For the Projected Landweber algorithm $2000$ iterations have
        been carried out.}
        \label{fig:ping_galaxy100_2}
\end{figure}
\begin{figure}
        \resizebox{\hsize}{!}{\includegraphics{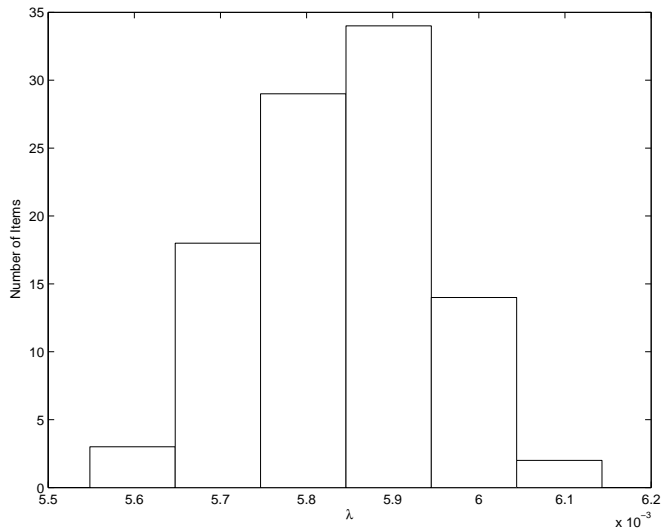}}
        \caption{Histogram of the values of the estimated parameter $\lambda$ for the $100$ experiments based on
	  Fig.~\ref{fig:ping_star_100_2}.}
        \label{fig:fig_hist}
\end{figure}
\begin{figure}
        \resizebox{\hsize}{!}{\includegraphics{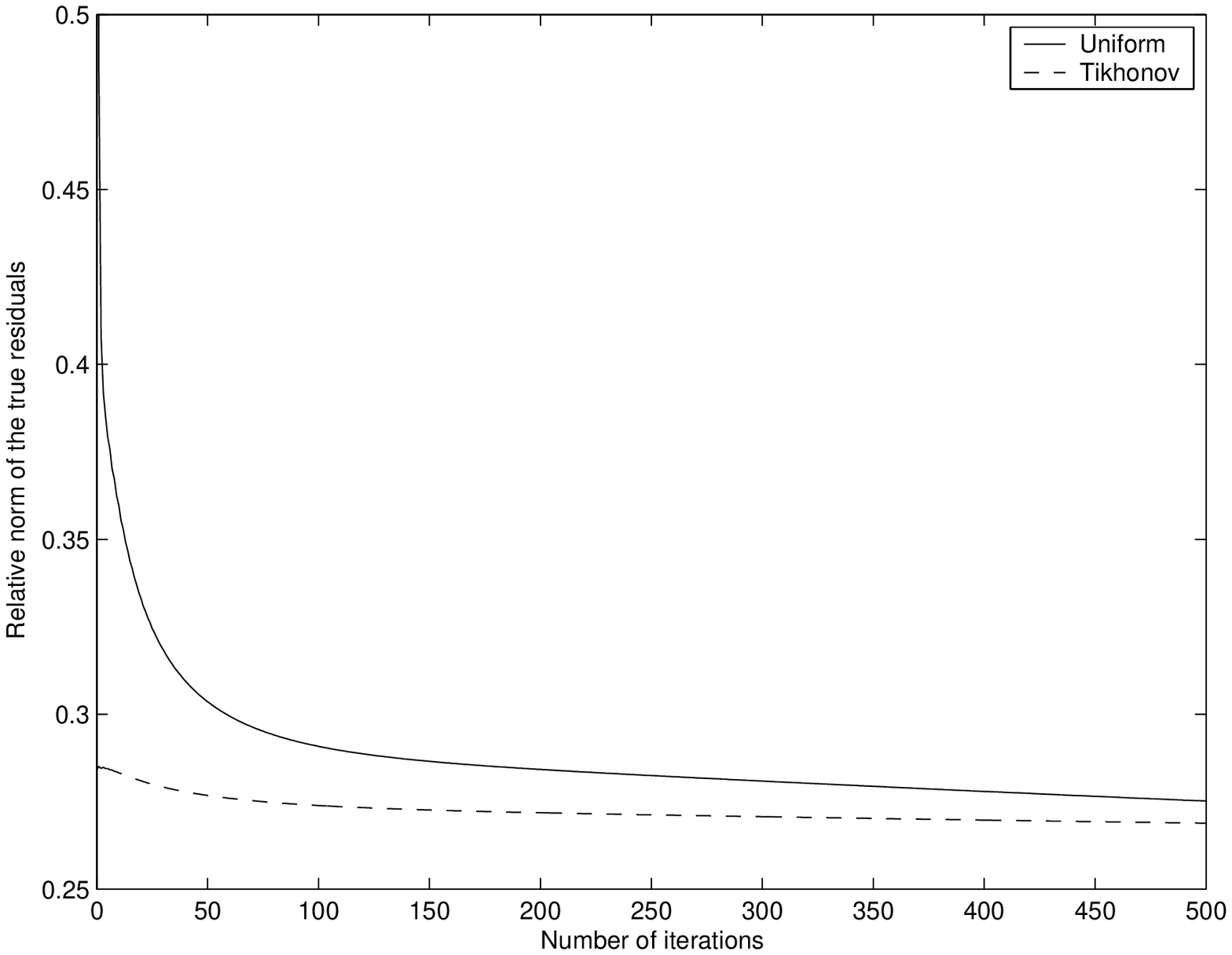}}
        \caption{Convergence rates for the Projected Landweber algorithm in the numerical experiment corresponding to
	  Fig.~\ref{fig:ping_sat_100_2}. Two starting guesses are used:
        the uniform image typical of the classic implementation of the algorithm, and the nonnegative
	  single-image Tikhonov shown in Fig.~\ref{fig:ping_sat_100_2}.}
        \label{fig:fig_iter}
\end{figure}

\clearpage

\appendix
\section{Derivation of Eq.\eqref{eq:gupdate}} \label{sec:gupdate}
Let $\eb_i$ be the vector with 1 in the $i$th entry and zero everwhere else, and $\fbt_{k,-i}$ the estimate
of $\fb$ that does not use the $i$th entry of $\gb_k$. We can write
\begin{equation}
g_{k,i} = \eb_i^T\gb_k,\qquad  \gh_{k,-i} = \eb_i^T\Ab_k\fbt_{k,-i}.\nonumber
\end{equation}
Let $\Pb_i$ be the projection that deletes the $i$th entry:
\begin{equation}
\gb_{k,-i} = \Pb_i\gb_k\nonumber
\end{equation}
so that $\Pb_i^T\Pb_i = \Ib - \eb_i\eb_i^T$. Let
\begin{equation}
\ab_{ki} = \Ab_k^T\eb_i,\qquad \Hb_k = \Ab_k(\,\Ab + \lambda\Ib\,)^{-1}\Ab_k^T,\nonumber
\end{equation}
then, by definition,
\begin{align}
 \fbt_{k,-i} & = \left(\,\sum_{i\neq k}\Ab_i^T\Ab_i + \Ab_k^T\Pb_i^T\Pb_i\Ab_k + \lambda\Ib\,\right)^{-1} \nonumber \\
&  \times \left(\,\sum_{i\neq k} \Ab_i^T\gb_i + \Ab_k^T\Pb_i^T\Pb_i\gb_k\,\right) \nonumber\\
= & \left(\,\Ab + \lambda \Ib - \Ab_k^T\eb_i\eb_i^T\Ab_k\,\right)^{-1}\left(\,\varthetab - \Ab_k^T\eb_i\eb_i^T\gb_k\,\right)\nonumber\\
= & \left[\,(\,\Ab + \lambda\Ib\,)^{-1} +
\frac{(\,\Ab + \lambda\Ib\,)^{-1}\ab_{ki}\ab_{ki}^T(\,\Ab + \lambda\Ib\,)^{-1}}
{1-\ab_{ki}^T(\,\Ab + \lambda\Ib\,)^{-1}\ab_{ki}}\,\right] \nonumber \\
& \times \left(\,\varthetab - \Ab_k^T\eb_i\eb_i^T\gb_k\,\right) \nonumber.
\end{align}
It follows that
\begin{align}
& \Ab_k\fbt_{k,-i} \nonumber =  \Ab_k\fbt - \Hb_k\eb_i\eb_i^T\gb_k \nonumber\\
& + \frac{\Hb_k\eb_i\ab_{ki}^T(\,\Ab + \lambda\Ib\,)^{-1}\varthetab - \Hb_k\eb_i\eb_i^T\Hb_k\eb_i\eb_i^T\gb_k}
{1-(\Hb_k)_{ii}}\nonumber
\end{align}
\begin{align}
\eb_i^T\Ab_k\fbt_{k,-i}  & - \gt_{ki} =  \frac{(\Hb_k)_{ii}}{1-(\Hb_k)_{ii}}\nonumber\\
& \times \left( \ab_{ki}^T(\,\Ab + \lambda\Ib\,)^{-1}\varthetab - \eb_i^T\gb_k\,\right) \nonumber \\
& = \frac{(\Hb_k)_{ii}}{1-(\Hb_k)_{ii}}\,
\left(\,\eb_i^T\Ab_k\fbh - g_{ki}\,\right) \nonumber\\
& = \frac{(\Hb_k)_{ii}}{1-(\Hb_k)_{ii}}\,\left(\,\gt_{ki}-g_{ki}\,\right) \nonumber
\end{align}
\begin{equation}
\gt_{k,-i} - g_{ki} = \gt_{ki} - g_{ki} + \frac{(\Hb_k)_{ii}}{1-(\Hb_k)_{ii}}\,\left(\,\gt_{ki}-g_{ki}\,\right)\nonumber
\end{equation}
\begin{equation}
\gt_{k,-i} - g_{ki} = \frac{1}{1-(\Hb_k)_{ii}}\,\left(\,\gt_{ki}-g_{ki}\,\right)\nonumber.
\end{equation}

\begin{acknowledgements}
We thank Prof. M. Bertero for a careful reading of the manuscript and his useful suggestions.
\end{acknowledgements}

\end{document}